\documentclass{elsart}
\usepackage{epsfig}
\newcommand{\ban}[1]{\begin{array}{#1}}
\newcommand{\ean}{\end{array}}

\newfont{\cf}{cmssi10 scaled \magstep5}
\newfont{\cc}{cmss10 scaled \magstep1}

\newcommand{\EE}[1]{ \mbox{$\times10^{#1}$} }

\newcommand{\eq}[1]{\mbox{Eq.~(\ref{#1})}}

\newcommand{\fig}[1]{\mbox{Fig.~\ref{#1}}}

\newcommand{\tab}[1]{\mbox{Tab.~\ref{#1}}}

\newcommand{\inv}[1]{\frac{1}{#1}}
\def\simleq{\; \raise0.3ex\hbox{$<$\kern-0.75em
      \raise-1.1ex\hbox{$\sim$}}\; }
\def\simgeq{\; \raise0.3ex\hbox{$>$\kern-0.75em
      \raise-1.1ex\hbox{$\sim$}}\; }

\def\atan{{\rm atan}}
\begin{document}
\begin{frontmatter}
\title{Wire scanners in low energy accelerators}
\author{P.~Elmfors\thanksref{PE1}},
\author{A.~Fass\`o\thanksref{AF1}},
\author{M.~Huhtinen},
\author{M.~Lindroos},
\author{J.~Olsfors} and
\author{U.~Raich}
\address{CERN, Geneva, Switzerland}
\thanks[PE1]{Present address: University of Stockholm, Fysikum, Box 6730, 
S-113 85 Stockholm, Sweden}
\thanks[AF1]{Present address: Stanford Linear Accelerator Center, 
Radiation Physics department, ms 48, P.O. Box 4349, Stanford CA 94309, USA}
\begin{abstract}
Fast wire scanners are today considered as part of standard instrumentation
in high energy synchrotrons. The extension of their use to synchrotrons 
working at lower energies, where Coulomb scattering can be important and
the transverse beam size is large, 
introduces new complications considering beam heating of the wire, 
composition of the secondary particle shower and geometrical consideration 
in the detection set-up. A major problem in treating these effects is that 
the creation of secondaries in a thin carbon wire by a energetic primary 
beam is difficult to describe in an analytical way. We are here 
presenting new results from a full Monte Carlo 
simulation of this process yielding information on heat deposited
in the wire, particle type and energy spectrum of secondaries and
angular dependence as a function of primary beam energy. The results
are used to derive limits for the use of wire scanners in low energy
accelerators.
\end{abstract} 
\end{frontmatter}
\newpage
\section{Introduction}
Fast wire scanners have been used successfully for beam profile 
measurements in the CERN-PS \cite{wr1,wr2,wr5b,wr3,wr5}
for more than ten years. We are presently considering to extend 
the use of them to the PS injector, the PS Booster. The PSB is a synchrotron
with four superimposed rings accelerating up to $1\EE{13}$  protons per ring 
from 50 MeV to 1 GeV kinetic beam energy. Operation of the fast wire scanners
in this low energy region where the physical beam emittance is large has 
triggered us to take a new look at the theory for 
wire heating, emittance blow up and the importance of the geometrical
relationships in the detector set-up. The work has been specially aimed at
the low energy domain of a proton beam but the results are in general 
valid also for the high energy domain.

\begin{figure}
\begin{center}
\mbox{\epsfig{file=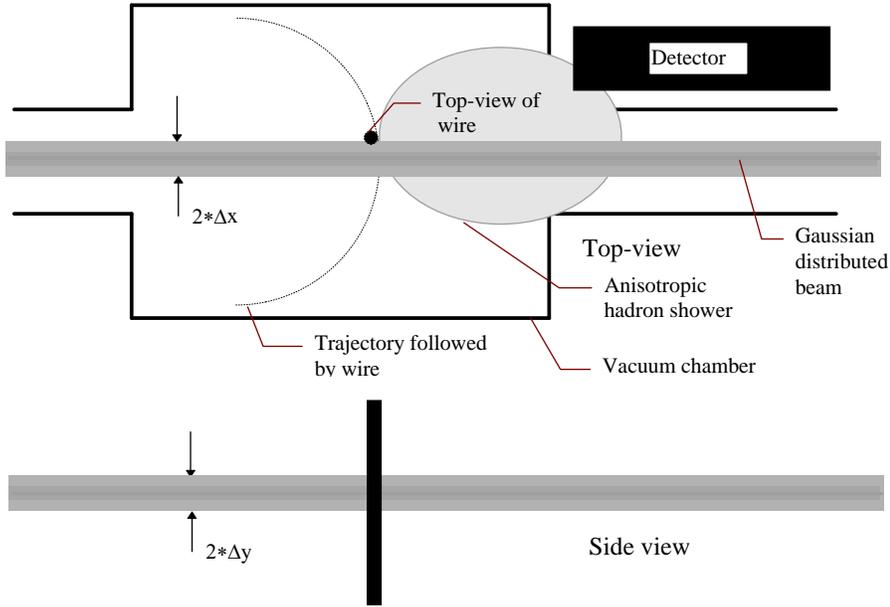,width=12cm,clip=}}
\end{center}
\caption{Geometry of beam and wire}
\label{f:profiles}
\end{figure}

The design and operation of fast wire scanners has been described elsewhere 
\cite{wr1,wr2,wr5b,wr3,wr5,wr5c,wr5d,wr4}. We will here only concern 
ourselves with the problems that set the limits for the use of these devices. 
For our discussion we need to define a geometry and as a starting point 
we use the geometry shown in   
\fig{f:profiles}. The figure is showing an instant in the process of the
wire sweeping through the beam. The transverse particle distribution within 
the beam is for simplicity taken to be Gaussian. As a measure for the beam size
we have in this work used $4 \times \sigma$ of the Gaussian beam profile as 
the measure 
for the beam width. This transforms to the so called $2\sigma$ emittance 
which at the CERN-PS is the standard quoted measure for transverse beam 
emittance.
%
\section{Simulation of beam-wire interaction}
The fast wire scanner method for measuring beam profiles is based on
the simple fact that an energetic particle beam passing any obstacle
, in our case a thin carbon wire, will create a secondary particle 
shower which is proportional to the primary beam intensity. The limits 
for the method is determined by how much deposited heat the carbon wire
can support and the possibility of detecting the secondary particles. With
this in mind we have simulated the process of a primary proton beam passing
a thin carbon wire using the FLUKA code, \cite{fluka}. The main interaction
parameters studied are the heat deposited in the carbon wire, the angular
dependence of the particle shower as a function of energy and the 
particle composition and energy spectrum of the secondaries.  
%
\subsection{Heat deposited in the carbon wire}
The simulation was especially aimed at 
calculating the part of the interaction energy deposited in the wire (which
eventually is transferred to heat). The fraction of the total deposited energy
leaving through the nuclear interaction proved to be very small, e.g. at 100
MeV kinetic beam energy 35.5 keV is deposited in wire as heat and only 
0.67 keV is leaving the wire through the nuclear interaction. The nuclear 
interaction part of the energy remains more or less constant up to the 
highest simulated value at 1 GeV. Furthermore, we also calculated the 
possible spread of the deposited energy along the wire due to internal 
scattering and even at the finest spatial resolution used in our 
simulations of 0.001 mm no significant smearing of the deposited energy 
was observed.   

As a model for the energy loss we take the Bethe--Bloch
formula (see e.g.\cite{Leo94})
\begin{equation}
\label{BB}
    \frac{dE}{dz}=2\pi N_a r_e^2 m_e c^2 \rho \frac ZA \frac{z^2}{\beta^2}
    \left[\ln\left(\frac{2m_ec^2 \beta^2 \gamma^2 W_{\rm max}}{I^2 }\right)
      -2\beta^2\right]~~,
\end{equation}
The symbols are explained in \tab{t:BB}. The maximum energy
transfer can be written as
\begin{equation}
\label{Wmax}
    W_{\rm max}=\frac{2m_e c^2\beta^2\gamma^2}{1+2s\gamma+s^2}~~.
\end{equation}
with $s=m_e/m_p$. The details of the notation is explained in e.g. 
\cite{Leo94}.
For protons in the beam $s$ is small and we can approximate
\begin{equation}
\label{Wmappr}
    W_{\rm max}\simeq 2 m_ec^2\beta^2\gamma^2~~,
\end{equation}
unless $2s\gamma\simgeq1$, i.e. the energy of the proton beam is 
more than about
2~TeV.

The values we use in
the Bethe--Bloch equation are compiled in \tab{t:BB} and the values wire
together with
\begin{table}
\caption{Numerical values used in the Bethe--Bloch equation for a carbon wire.}
\begin{tabular}{llll}
   \hline
   $N_a$  &Avogadros Number & $6.022\EE{26}$  & $\EE{-3}$mol$^{-1}$\\
   $r_e$  &classical electron radius & $2.817\EE{-15}$ & m  \\
   $m_e$  &electron rest mass & $9.109\EE{-31}$ & kg  \\
   $c$    &light velocity & $2.9979\EE{8}$  & m\,s$^{-1}$  \\
   $\rho$ &graphite density & $1.77\EE{3}$    & kg\,m$^{-3}$ \\
   $Z$    &atomic number & 6                  &   \\
   $A$    &atomic weight & 12                 &   \\
   $z$    &charge of incident particle & 1                  &   \\
   $W_{\rm max}$&maximum energy transfer & $1.637\EE{-13}\beta^2\gamma^2$ & kg\,m$^2$\,s$^{-2}$  \\
   $m_p$  &mass of incident particle & $1.672\EE{-27}$ & kg  \\
   $I$    &mean excitation potential & $1.266\EE{-17}$ & kg\,m$^2$\,s$^{-2}$  \\
   \hline
\end{tabular}
\label{t:BB}
\end{table}
\begin{figure}
\unitlength=1mm
\begin{picture}(90,90)(0,0)
\includegraphics{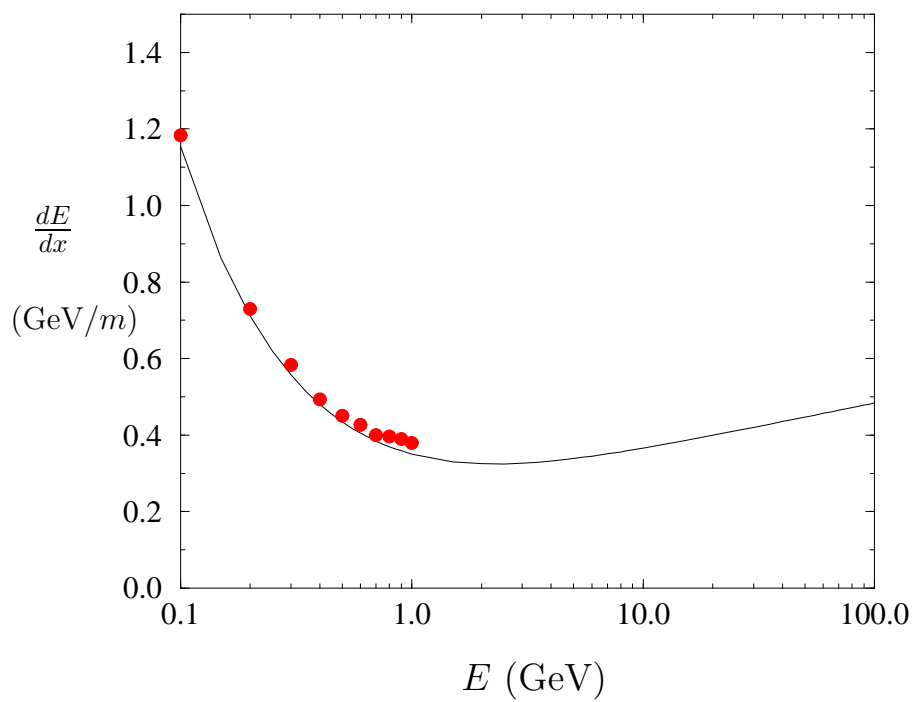}
   \put(0,0){}
   \put(72,0){\large $E$ (GeV)}
   \put(15,60){\large $\frac{dE}{dx}$}
   \put(12,48){(GeV/$m$)}
\end{picture}
   \caption{Simulated and Bethe--Bloch values of $dE/dz$ for a carbon
     wire as a function of kinetic energy 
     , $E=m_p c^2 (\gamma-1)$.}
   \label{f:dEdx}
\end{figure}
other numerical values are given in \tab{t:phval}.
\begin{table}
\caption{Additional accelerator, beam and wire parameters}
\begin{tabular}{llll}
   \hline
   $\eta$ &emissivity (\rm{C}) & $0.88$  & \\
   $\sigma$ &Stephan--Boltzmann constant & $5.67\EE{-8}$  & kg\,s$^{-3}$\,K$^{-4}$\\
   $c_V$ &heat capacity (\rm{C})& $1.25\EE{6}$  & kg\,s$^{-2}$\,m$^{-1}$\,K$^{-1}$\\
   $r$ &wire radius & $1.5\EE{-5}$  & m\\
   $N$ &number of beam particles & $2\EE{13}$  & \\
   $\tau_0$ &revolution time at $\beta=1$ & $2.1\EE{-6}$  & s \\
   $\epsilon_{Nx}$ &normalised beam emittance, x & $1.7\EE{-4}$  & m \\
   $\epsilon_{Ny}$ &normalised beam emittance, y & $9.0\EE{-5}$  & m \\
   $\beta_{Tx}$ &Twiss value, x & $12$  & m \\
   $\beta_{Ty}$ &Twiss value, y & $21$  & m \\
   $v$ &wire velocity & $20$ & m\,s$^{-1}$ \\
   $\kappa$ &heat conductivity (\rm{C})& $150$  & W\,m$^{-1}$K$^{-1}$ \\
   \hline
\end{tabular}
\label{t:phval}
\end{table}

  In \fig{f:dEdx} we compare the Bethe--Bloch equation with the 
Monte--Carlo simulation. The deviation between the theory and the
simulation is small and justifies using the Bethe--Bloch equation for the
rest of the analysis.
%
\subsection{Secondary particle shower}
%
The inelastic cross section of a 100\,MeV proton in carbon is of the
order of 240\,mb, rising to about 250\,mb at 1\,GeV. The elastic cross 
section drops from about 180\,mb at 100\,MeV to roughly 100\,mb at 1\,GeV.

For the typical total scattering cross section of 400 mb and a carbon density 
of 2.3g/cm$^3$ the mean free path in the wire material is 21.7\,cm. 
For a wire with a radius of 15\,$\mu$m
the average number of protons needed to obtain 
one scattering is 9200. Since the interaction probability is so small, 
scattering of the produced secondaries can be neglected.

Using the scattered particles the beam intensity can be monitored, by 
measuring the energy deposition in a detector well away from the beam axis. 
This requires that the energy deposited per one scattered particle 
is known. This can be calculated with the FLUKA Monte Carlo program 
\cite{fluka}.

The simulations should be carried out in a realistic geometry, 
since stray radiation around an accelerator may give a significant 
contribution to the total energy deposition. The simulations 
presented in the following were carried out in a very idealized geometry.

The detectors were represented by polyethylene disks of 5\,mm thickness, 
 3\,cm diameter and a density of 0.95\,g/cm$^3$. These were placed at a 
radial distance of 50\,cm from the point where the beam hits the wire. 17 
detectors, starting at a polar angle of 10 degrees and spaced by 10 degrees 
were used.
									 
The scattering of protons with 100\,MeV and 1\,GeV kinetic energy was 
studied. Instead of simulating a beam hitting the wire, we first created sets
of 10000 elastic and 10000 inelastic events for both energies.
The secondaries from these events were then mixed in ratios given by the 
inelastic and elastic cross sections. Since these event sets were postulated 
to be representative of an infinite number of events, the event structure 
itself was not important. So the azimuth angle of each secondary particle 
could be separately sampled between zero and 2$\pi$. Due to the small 
solid angle covered 
by the detectors the relatively limited number of events could be reused 
several times to improve the statistics of the quantities scored at the
detectors. 

\begin{figure} 
\setlength{\unitlength}{1cm}
 \begin{picture}(10,12)
   \put(0.0,0.0){\mbox{\includegraphics[height=12cm]{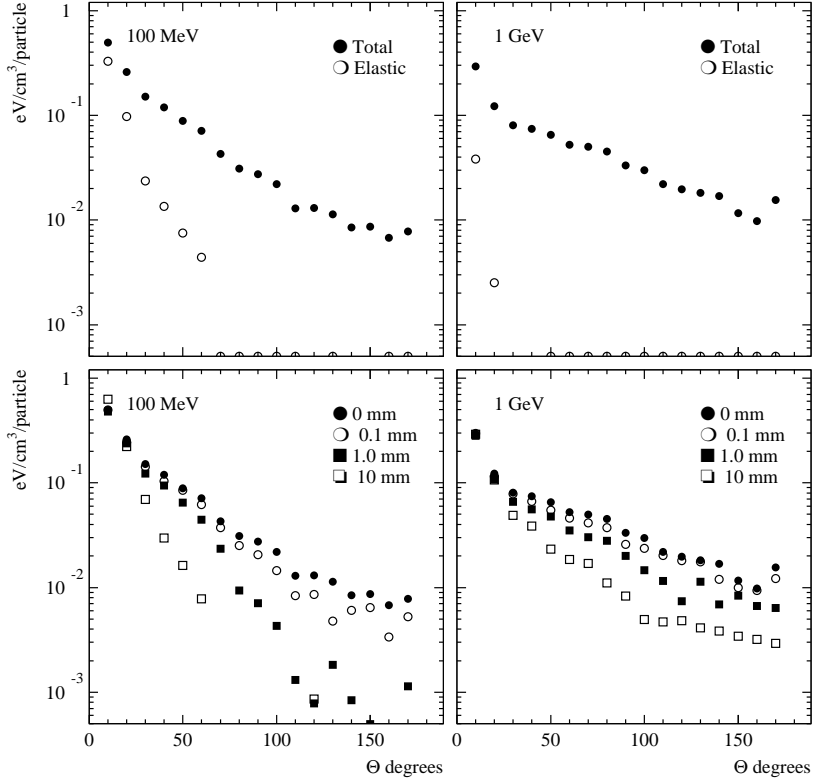}}}
 \end{picture} 
\caption{Energy deposition in the polyethylene detectors as a function of 
angle with respect to the beam direction. Normalisation is to one particle 
incident on the 30\,$\mu$m wire.}
\label{edep}
\end{figure}

The obtained energy deposition is shown in \fig{edep}. Normalization 
is to one proton incident on the wire assuming a circular wire 
with 30\,$\mu$m diameter. In the upper plots the total energy deposition and 
the fraction coming from elastic scattering is shown separately. It can be 
seen that the elastic contribution quickly becomes negligible, which is 
fortunate, since the angular distributions of the elastic  scattering in 
{\sc Fluka} are not really optimized to reproduce single scattering 
distributions. In particular they lack the diffractive structure, which is 
characteristic for elastic scattering and only reproduce rough trends of the 
differential cross section.

\begin{figure} 
\setlength{\unitlength}{1cm}
 \begin{picture}(10,12)
   \put(0.0,0.0){\mbox{\includegraphics[height=12cm]{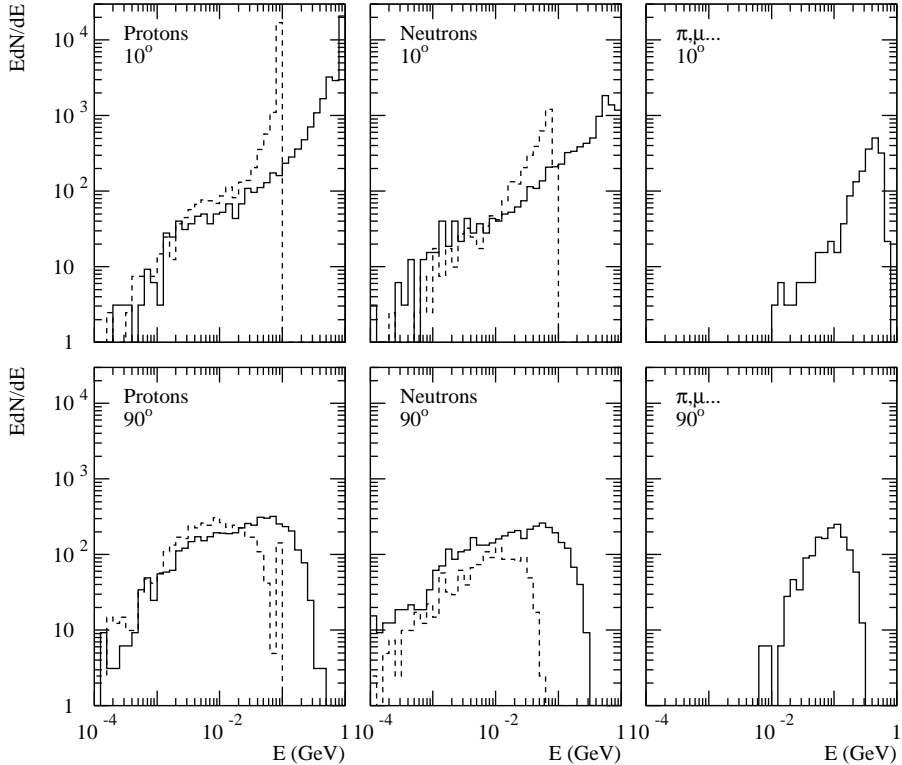}}}
 \end{picture} 
\caption{Kinetic energy spectra of secondaries, the solid line shows the 
spectrum for a primary proton beam of 1 GeV kinetic energy and the dashed
line for a primary proton beam of 100 MeV. Note that direct photons 
are not an important contribution and have not been plotted. The neutral 
pions however are included in the plots of the third column.}
							  
\label{f:spectra}
\end{figure}

The lower plots show a comparison of the energy deposition in the detectors 
if material is introduced between the wire and the detectors. In a real 
situation there will always be at least a thin window, sometimes a thick 
beam pipe. Therefore we compare four cases, our idealized one without any 
material and with three steel window thicknesses of 0.1, 1 and 10\,mm. This 
``window'' was in the simulations a spherical iron (density 7.87\,g/cm$^3$) 
shell of 40\,cm radius, surrounding the point where the particles 
originated from. It can be seen that the thicker windows start to reduce 
the energy deposition only at large angles. At the forward angles the 
presence of material can even increase the energy deposition due to 
secondary interactions. As expected the material has more effect for the 
lower beam energy. Qualitatively the behaviour can be understood by looking 
at the particle spectra. These are shown in \fig{f:spectra} for 
scattering angles of 10 ($\pm$ 5) degrees and 90 ($\pm$ 5) degrees.
It can be seen that at 90 degrees the particle spectra are considerably 
softer than in the forward direction, which accounts for the larger effect 
of material. Of course also the spectra for the lower beam energy are softer 
than those from the higher one. A minimum ionizing particle would loose 
about 12 MeV/cm in iron, which already would cut into the spectrum. In fact
all of the particles are on the 1/$\beta^2$ part of the Bethe--Bloch equation, 
and therefore lose much more than 12 MeV/cm. And with decreasing energy this 
energy loss increases rapidly. So a 10\,mm iron layer stops a significant 
fraction of the particles.

Two aspects should be kept in mind when interpreting these results:
\begin{enumerate}
\item At the low energies considered here evaporation fragments and splitting 
of the $^{12}$C nucleus into three helium nuclei are important inelastic 
channels. The heavy fragments have not been transported. This underestimates 
the result and probably their effect would be to make the angular distribution
flatter, since elastic scattering and particle production are less isotropic 
than evaporation and fragmentation. However, the heavy fragments are very 
slow and therefore highly ionising, so they most probably would be stopped in 
almost any kind of window separating the wire from the detectors.
\item As was pointed out previously the simulations should be done in a 
realistic geometry with realistic beam halo. Neither surrounding walls, nor 
support materials, nor the arriving beam have been included in the 
simulations. Their effect would be to generate stray radiation, mostly 
photons and neutrons all over the system. Especially plastic scintillators 
would be sensitive to this stray radiation field.
\end{enumerate}	
%
\section{Beam heating of the wire}
\subsection{A simple model of heating}
In the process of is interaction with the beam the wire the wire gets heated 
by energy absorption. We shall formulate a model to estimate the maximal 
temperature rise in the stationary situation when the wire sweeps back
and forth with a given frequency. The formalism is generally valid for 
any particle type and any energy range. The numerical examples are 
given for a low energy proton
beam. For a general discussion on heating in any obstacle by a primary beam
we refer to \cite{warner}, for a detailed discussion of the heating in a
carbon wire by protons to
\cite{elm97} and for high energy electrons to \cite{wr6}. We compile the
main results here for convenience. Let us initially
neglect conductive cooling and only consider radiative cooling, in the
end of this section we will derive the condition for this 
approximation. 

When the wire is in the beam it is thus heated according
to the equation
\begin{equation}
\label{heateq}
    \frac{dT}{dt}=\frac{N\beta}{A\tau_0c_V}\frac{dE}{dx}
    -\frac{2\sigma\eta T^4}{r c_V}
    \equiv a-bT^4~~,
\end{equation}
and the maximal temperature it can reach is
\begin{equation}
\label{Tm}
    T_m=\left(\frac{N\beta r}{2A\tau_0\sigma\eta}\frac{dE}{dx}\right)^{1/4}~~.
\end{equation}
For constant normalised emittance it is 
useful re-express everything in terms
of $\beta$. Furthermore, we must also account for the fact that the highest 
temperature is reached in the part of the wire that sweeps through 
the centre of the assumed Gaussian beam profile. So all together we
can write the maximum temperature as,
\begin{equation}
\label{Tmphys}
    T_m^4=\frac{4Nr}{4\pi\,\tau_0\sigma\eta}\,
    \inv{\sqrt{\epsilon_{Nx}\epsilon_{Ny}\beta_{Tx}\beta_{Ty}}}\,
    \frac{\beta^2}{\sqrt{1-\beta^2}}\,
    \frac{dE}{dz}~~.
\end{equation}
From the previous section we know that the energy deposited as heat in the
wire is well described by the Bethe--Bloch formula. The 
asymptotic form of $dE/dz$ for two ranges of $\beta$ can be written
as
\begin{eqnarray}
\label{dEasymp}
    \frac{dE}{dz}&\sim&\inv{\beta^2}(\ln\beta+{\rm const.})~~;
    \quad{\rm for}~\beta\ll 1~~,\\
    \frac{dE}{dz}&\sim&\ln\inv{1-\beta^2}~~;
    \quad{\rm for}~1-\beta^2\ll 1~~.
\end{eqnarray}
We assume that $\beta$ is not so small that $dE/dz$ 
changes sign. 
The consequences for $T_m$ in the limiting cases are 
\begin{eqnarray}
\label{Tmasymnp}
    T_m&\sim&\ln\beta+{\rm const.}~~;
    \quad{\rm for}~\beta\ll 1~~,\\
    T_m&\sim&(1-\beta^2)^{-1/8}\left(\ln\inv{1-\beta^2}\right)^{1/4}~~;
    \quad{\rm for}~1-\beta^2\ll 1~~.
\end{eqnarray}
That is, for small $\beta$ it 
goes to a constant up to logarithmic corrections, while at $\beta\simleq 1$
the maximal temperature increases with the beam energy $E$ 
like $T_m\sim E^{1/4}$. This estimate
is valid for a wire which remains in the centre of the beam. 
If the wire is swept with
a constant speed and frequency the time the wire spends in the beam
decreases with increasing $E$ since the beam size decreases, thus seemingly 
reducing the
final temperature. In order to find out what really happens in that case we
need to solve the dynamical heating problem.
%
\subsection{Solution to the periodic heating}
We shall now find
the stationary temperature in the case the wire is swept through the
beam with a given speed and frequency. The wire is cooled by radiation
from a temperature $T_0$ to $T_1$ during the cooling time $t_c$, and then
heated to $T_2$ during the heating time $t_h$. Putting $T_0=T_2$ we
shall find the maximal temperature in the stationary situation.
During cooling the temperature is governed by the equation
\begin{equation}
\label{cooling}
   \frac{dT}{dt}=-bT^4~~,
\end{equation}
with the solution
\begin{equation}
\label{coolsol}
    T_1=T_0\left(\inv{1+3bt_cT_0^3}\right)^{1/3}\equiv T_0 \alpha(T_0)~~.
\end{equation}
We can interpret 
$t_h$ as an effective heating time
taking into account the variation of the beam intensity. After solving
 \eq{heateq} and equating
$T_0$ and $T_2$ we find the implicit equation for $T_0$
\begin{equation}
\label{T0eq}
   2\left(\atan\frac{T_0}{T_m}-\atan\frac{T_0\alpha}{T_m}\right)
   +\ln\left[\frac{(T_m+T_0)(T_m-T_0\alpha)}{(T_m-T_0)(T_m+T_0\alpha)}
   \right]=4bt_hT_m^3~~.
\end{equation}
We can gain some insight by solving this equation in the limiting cases of
very long and very short cooling times, i.e. $\alpha\simeq0$ and 
$\alpha\simeq1$.
\subsubsection{Long cooling time}
When the cooling time is long,
\begin{equation}
\label{tclong}
    t_c\gg\frac{rc_V}{6\sigma\eta T_0^3}~~,
\end{equation}
i.e. $\alpha(T_0)\simeq0$, \eq{T0eq} reduces to
\begin{equation}
\label{T0tclong}
    2\,\atan\frac{T_0}{T_m}+\ln\frac{T_m+T_0}{T_m-T_0}=4bt_h T_m^3~~,
\end{equation}
which has the approximate solution
\begin{eqnarray}
\label{thlsol}
   T_0&\simeq&T_m~~, \quad bt_hT_m^3\simgeq1~~,\\
   T_0&\simeq&bt_hT_m^4~~, \quad bt_hT_m^3\simleq1~~.
\end{eqnarray}
Using the effective heating time
\begin{equation}
\label{th}
    t_h=\frac{2\Delta x}{v}=\frac{\sqrt{\epsilon_{Nx}\beta_{Tx}}}{v}
    \left(\frac{1-\beta^2}{\beta^2}\right)^{1/4}~~,
\end{equation}
we have in the limit of large and small beam energy
\begin{eqnarray}
\label{tclasymp}
    T_0\simeq T_m\sim{\rm const.}&&{\rm since}~bt_hT_m^3\rightarrow\infty~~{\rm as}~
    \beta\rightarrow0~~,\\
    T_0\sim (1-\beta^2)^{1/8}\left(\ln\inv{1-\beta^2}\right)^{1/4}
    &&{\rm since}~bt_hT_m^3\rightarrow0~~{\rm as}~
    \beta\rightarrow1~~.
\end{eqnarray} 
In the case the final temperature is small it is necessary to pay extra 
attention to the condition in \eq{tclong} since it tends to be more
difficult to satisfy.
\subsubsection{Short cooling time}
It is also possible to find an approximative solution to \eq{T0eq}
in the case the cooling time is very short. Then $\alpha\simeq1$ and we
also expect that $T_0\simeq T_m$. Doing an expansion in small deviations
we find
\begin{equation}
\label{T0tcs}
    T_0\simeq T_m\left(1-\frac{2t_c}{rc_V}
      \frac{\sigma\eta T_m^3}{\exp[\frac{8\sigma\eta T_m^3 t_h}{rc_V}]-1}
      \right)~~,
\end{equation}
but, as we shall see, the cooling time has to be very short for this
approximation to be valid.
\subsubsection{Conductivity}
So far we have neglected conductivity in the wire and we shall now estimate
its importance compared to radiation. 
As a measure of the importance of conductivity we shall compare the
radiated energy from the heated region of the wire with the conducted
energy. The conducted energy per unit time is
\begin{equation}
\label{Pcond}
    P_c=-\pi r^2\kappa\frac{\partial T}{\partial y}(0)=
    2\pi r T_0^{5/2}\left(\frac{\eta\sigma\kappa r}{5}\right)^{1/2}~~,
\end{equation}
where
\begin{equation}
\label{Tstat}
    T(y)=T_0\left(1+\frac{y}{L}\right)^{-2/3}~~,\quad 
    L=\left(\frac{5r\kappa}{9\eta\sigma T_0^3}\right)^{1/2}~~.
\end{equation}
while the radiated power is
\begin{equation}
\label{Prad}
    P_r=4\pi r \Delta y\,\eta \sigma T_0^4~~.
\end{equation}
The conductivity can be neglected when the condition
\begin{equation}
\label{prpccond}
    \frac{P_r}{P_c}=\frac{10\Delta y}{3L}
    =\left(\frac{5\eta\sigma T_0^3\epsilon_{Ny}\beta_{Ty}}
      {r\kappa\beta\gamma}\right)^{1/2}\gg1~~,
\end{equation}
is satisfied.
%
\subsection{Numerical example}
In order to better see the validity of the approximations and the actual
physical values they predict we shall go through a real
example with a carbon wire in a proton beam. It is important to 
remember that the nature of our problem and the many approximations
done in our derivation are such that we
can't hope for high numerical precision results. The derived formulas
can only give us an idea of the temperatures reached and show 
the temperatures energy dependence. The values we use in
the Bethe--Bloch equation are given in \tab{t:BB}. 
Other numerical values are given in \tab{t:phval}.

\begin{figure}
\unitlength=1mm
\begin{picture}(90,90)(0,0)
\includegraphics{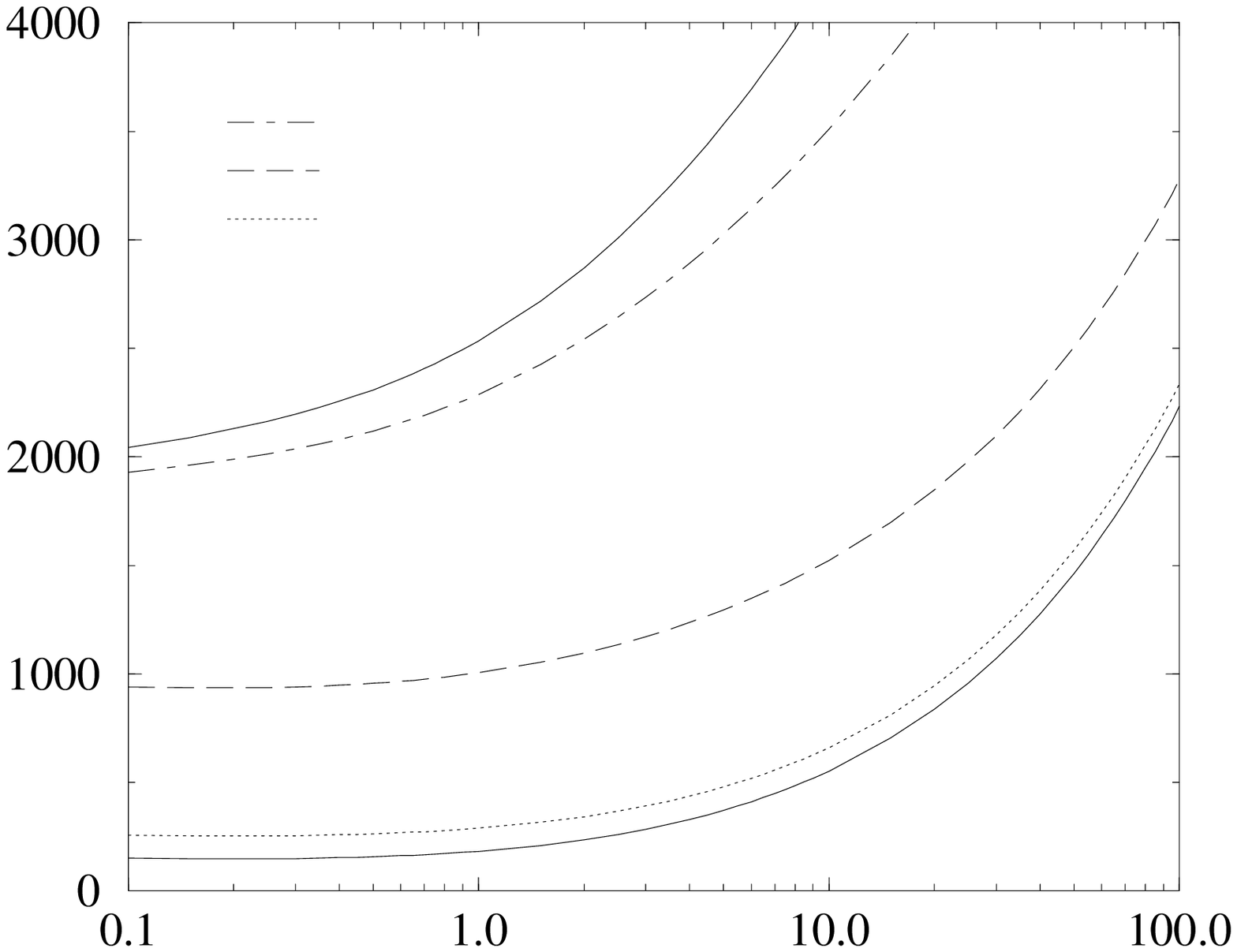}
   \put(0,0){}
   \put(72,5){$E$ (GeV)}
   \put(15,60){$T_0$ (K)}
   \put(60,50){\small $T_m$}
   \put(55,80){\small $t_c=0.001$ s}
   \put(55,76){\small $t_c=0.1$ s}
   \put(55,72){\small $t_c=100$ s}
   \put(107,18){\small $T_0$, $t_c=\infty$}
\end{picture}
   \caption{Exact solution to \eq{T0eq} for several values of $t_c$.
     The upper solid line indicates $T_m$ and the lower one is
     $T_0$ in the long cooling time approximation.}
\label{f:T0}
\end{figure}

\begin{figure}
\unitlength=1mm
\begin{picture}(90,90)(0,0)
\includegraphics{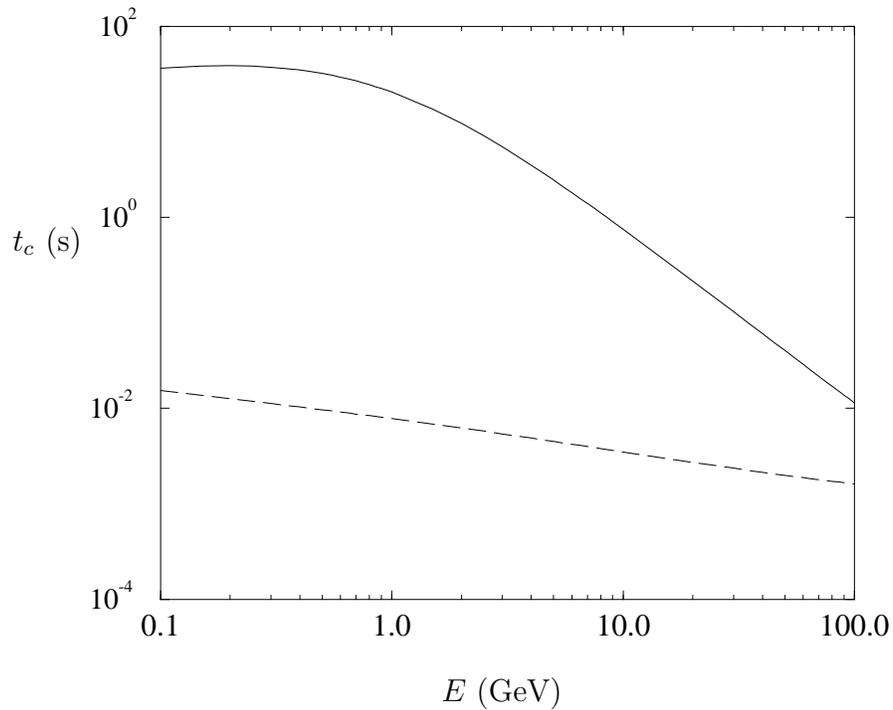}
   \put(0,0){}
   \put(72,0){$E$ (GeV)}
   \put(15,60){$t_c$ (s)}
\end{picture}
\caption{The long cooling time approximation is valid when $t_c$ is much
     larger than the values shown by the solid line above, while
     the short cooling time is valid far below the dashed line.
     There is thus a large interval where none of these approximations
     is valid.}
\label{f:tc}
\end{figure}

With the parameters in the tables the maximal temperature is given by
\begin{equation}
\label{Tm4num}
    T_m=1200\,(1-\beta^2)^{-1/8}
    \left[\ln\frac{\beta^2}{1-\beta^2}+9.47-\beta^2\right]^{1/4}~{\rm K}~~.
\end{equation}
In \fig{f:T0} the upper solid line shows $T_m$ as a function of beam energy 
$E$ using
\begin{equation}
\label{bE}
    \beta(E)=\sqrt{1-\left(\frac{m_pc^2}{E+m_pc^2}\right)^2}~~.
\end{equation}
For long cooling time we can use the approximate formula in \eq{thlsol}.
It turns out that the combination $bt_hT_m^3$ is in fact small for a large
range of energies so the approximate temperature is $T_0=bt_hT_m^4$.
We show this as a function of energy as the lower solid line in \fig{f:T0}.
The condition for the long cooling time to be valid is 
from \eq{tclong} that
\begin{equation}
\label{tccondnum}
    t_c\gg\frac{6\EE{7}~{\rm K}^3}{T_0^3}~{\rm s}~~.
\end{equation}
The right hand side of this equation is plotted 
as the solid line in \fig{f:tc}.

For low energies the sweep frequency must be well below 0.01 Hertz for
this approximation to be valid.
The approximation of short cooling time is valid if the correction in
\eq{T0tcs} is small, which means
\begin{equation}
\label{Ttcscond}
    t_c\ll\frac{rc_V}{2\sigma\eta T_m^3}
    (\exp[\frac{8\sigma\eta T_m^3 t_h}{rc_V}]-1)~~,
\end{equation}
and this limit is also plotted in \fig{f:tc} (dashed line).  
Since there is a wide range of cooling times for which none of the above 
approximations works we should also solve the full \eq{T0eq} exactly.
This solution is presented in \fig{f:T0} for $t_c=0.001$, $0.1$ and
100~s.

Finally we need to check from \eq{prpccond} 
that the conductivity is negligible in our
example. At high energy $T_0$ increases but at the same time $\Delta y$
decreases, and the net effect is that the ratio $P_r/P_c$ decreases.
For the region in $E$ that we have studied here we always have $P_r/P_c>10$
which justifies neglecting conductivity.
%
\section{Emittance blow-up of the primary beam}
%
\subsection{Emittance blow-up due to a thin window}
%
The increase of the beam emittance when passing a thin window  is a well 
understood process (see e.g. \cite{CAS92I}). 
The scattering of the beam increases the angular 
spread of the beam which 
through filamentation results in an increased emittance
\begin{equation}
\label{deltaE}
E=E_0+\Delta E=E_0+\frac{\pi}{2} \beta _{T} \langle\theta ^2\rangle. 
\end{equation}
Here $E_0$ is the initial emittance and $\beta_{T}$ the Twiss value in 
the plane of the emittance at the thin window. 
The average square scattering angle will depend on the characteristics of the 
foil and the beam and is usually 
derived using formulas based on the Moli\`{e}re theory for multiple Coulomb 
scattering \cite{bet89}. The 
emittance blow up due to the wire scanner device can be evaluated using the 
same formalism as the wire can
be pictured as a virtual foil which thickness depends on the velocity 
and shape of the wire and the 
velocity of the beam. For the case  of a  wire with a circular cross section 
in a synchrotron with a revolution 
time of $\tau_0$ (at $\beta=1$) the virtual foil thickness (vft) can be 
written as
\begin{equation}
\label{f:thick}
z_{vft} = \frac{(2r)^2\pi\beta}{4 v\tau_0}~~.
\end{equation}
\subsection{Scattering theory}
For small deflection angles a good approximation 
for the average root mean square scattering angle is given by 
\cite{prd94a,prd94b}
\begin{equation}
\label{mol}
    \theta _0=\frac{13.6{\rm \;MeV}}{\beta pc}Q~\sqrt{\frac{z}{X_0}}
      \left(1+0.038 \ln\left(\frac{z}{X_0}\right)\right),
\end{equation}
where $p, \beta c$ and $Q$ are momentum, velocity and charge number 
of the incident particles and $\frac{z}{X_0}$ is 
the thickness of the scattering medium in radiation lengths ($z$ being the 
coordinate along the beam-line). However, 
the formula is only accurate to about 11\%  
or better for $1\EE{-3}<\frac{z}{X_0}<100$. For a  typical wire scanner
with $z_{vft}$ according to 
\eq{f:thick}, $\frac{z_{vft}}{X_0}$ is much smaller than  $1\EE{-3}$.
Consequently, we are for e.g. the CERN-CPS wire scanners left in a 
situation in between single Coulomb scattering and multiple 
scattering. We can get an idea of the order of magnitude if we assume that 
we are dealing with  single Rutherford scattering events and that 
outside the atomic radius there is no 
interaction between the primary particle and the scattering centre. If for 
a numerical example we take the  
parameters in \tab{t:phval} at a kinetic proton beam energy of 1 GeV  this 
approach gives a root mean square 
scattering angle of typically a few $1\EE{-11}$ radians while a multiple
scattering approach using \eq{mol} yields 
a few $1\EE{-7}$ radians. The large range spanned by these two extreme 
approaches might be of theoretical interest but of no practical importance in 
a large physical beam emittance machine where both values yield an emittance
blow-up well below the 
required  precision. An attempt was done to measure 
the emittance blow-up in the CERN-CPS caused by the passage of a scanner wire
using a two sweep process on a 500 ms
flat-top of the magnetic cycle. An initial sweep and measurement was 
followed approximately 400 ms later by a back-sweep also with a measurement.
The difference in emittance between the first and the second sweep was assumed
to mainly be due to the blow-up in the wire. Earlier experience at 
the CPS has shown that this is a reasonable assumption. However, for a physical
$\epsilon_h=30\pi$ mm mrad proton beam at a beam energy of 
300 MeV and a wire velocity of 20 m/s the {\it error} of the measured 
``emittance blow-up'' was  $0.3\pi$ mm mrad which is insuficient to separate 
between the two discussed approximations for the scattering process.
Planned improvements, in line with findings presented in this note, for the 
dedicated low energy wire scanners in the PSB booster should make that 
possible.
%
\section{Detection of secondary particles}
In section 2 we have shown that a detectable amount of secondary 
particles are created by 
the primary beam when passing the wire of the fast wire scanners. 
A detector positioned at a given polar angle with respect to the
beam direction can be used to monitor the number of particles
scattered by the wire. If the angle is of the order of 10 degrees
or larger, we know from section 2 that only nuclear scattering events, 
both elastic and inelastic, contribute.
%
\subsection{Effect on the deduced emittance}
%
%
\begin{figure}
\begin{center}
\epsfig{file=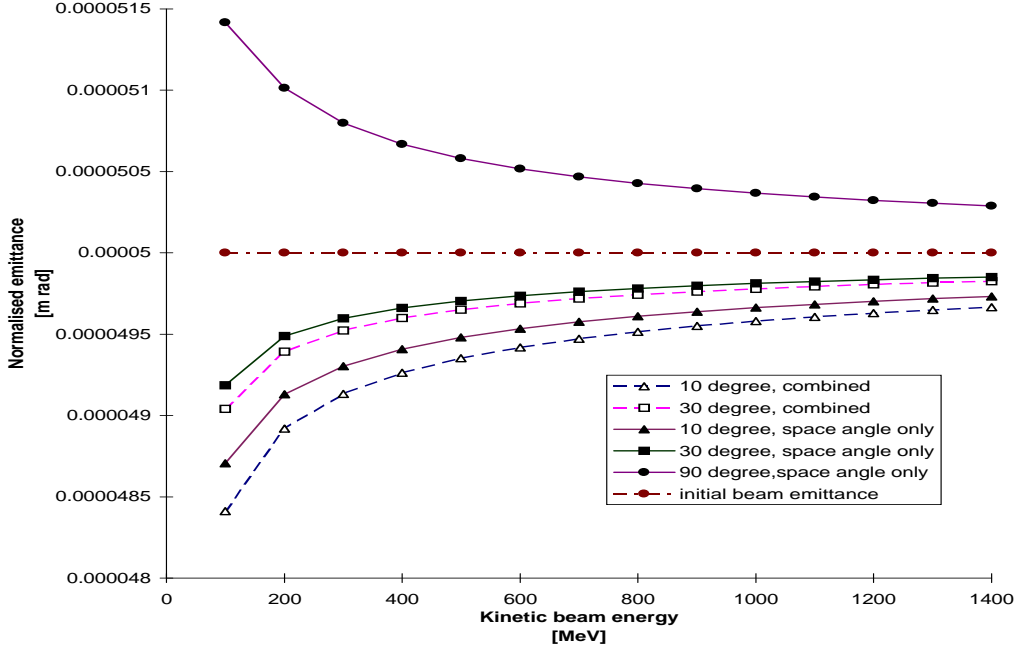,height=10cm,width=14cm,clip=}
  \caption{The deduced emittance is deviating from the real beam 
           emittance due to geometrical effects and the anisotropy
           of the induced particle shower. The effect is increasing
           with decreasing energy (larger transversal beam size). Both
           the combined result for both effects and the more significant
           geometrical effect are plotted for 10,~30 and 90 degree (only
           geometrical effects at 90 degree as the anisotropy effect 
           of \eq{wtheta} is zero at 90 degree).}
\label{f:anis}
\end{center}
\end{figure}
The detectors occupy a certain space-angle which together with other 
detector specific parameters 
determines the detection efficiency. For a beam which is large transversely 
such as the high intensity - 
low energy proton beam in the CPS, the space angle will change noticeably 
as the wire is passing through 
the beam.

The secondary particles will be emitted anisotropically with a majority
of the particles going in the forward 
direction. This anisotropy will have ``skewing'' effect on the measured 
beam profile if the transverse 
beam size is large. For a rough estimate of the effect we will assume that 
the anisotropy is described by
\begin{equation}
\label{wtheta}
     W(\theta)=1+\cos\theta
\end{equation}
A simple approach to calculate the size of the combined effect is to 
divide the beam into thin slices, 
calculate the detector efficiency and the resulting number of detected 
particles for each slice and finally 
compare the initial emittance with the deduced emittance. Using the 
wire scanner example from  \tab{t:phval}
we have calculated the influence of the change in space angle and 
of the particle shower anisotropy for two wire scanner set-ups. 
In the first configuration the detector is positioned in the forward 
direction 
15 cm from the wire at an angle of 
30 degree to the beam axes and in the second configuration with an 
angle of 10 degree to the beam axes. The configuration with an angle 
of 30 degree to the beam axes is in the PS-complex enforced by the 
space limitations at the 
wire-scanner installations. In our numerical example the beam was initially 
assumed to be Gaussian and
$\sigma$ for the measured beam profile was calculated as

\begin{equation}
        \sigma ^2 = \frac{\sum\limits_{x=\rm first~channel}^{\rm last~channel}
        {(x_i-\overline{x})^2Ch(x_i)}}
        {\sum\limits_{\rm last~channel}^{\rm first~channel}{Ch(x_i)}}~~,
\end{equation}
where

\begin{equation}
        \overline{x}=\frac{\sum\limits_{x=\rm first~channel}^{\rm
        last~channel}
        {x_iCh(x_i)}}{\sum\limits_{\rm last~channel}^{\rm first~channel}
        {Ch(x_i)}}~~.
\end{equation}
and where $Ch(x_i)$ is the number of counts in channel $x_i$.

In \fig{f:anis} we can see that the deviation form the real beam emittance 
is, as expected, increasing with
decreasing beam energy. That is to say increasing with increasing
transverse beam size. In \fig{f:profile} the measured beam profile during 
acceleration of the same beam at
100 MeV and 1.4 GeV are shown for a detector positioned at 85 degree angle
to the beam axes. The beam is transversally larger at 100 MeV and the beam
profile is slightly deformed due to the discussed geometrical effects. The 
resulting error in the deduced normalised
emittance is small and will in most situations 
be insignificant. However, it is interesting to note that for the Gaussian 
beam shape the deformation of the beam profile is such that the influence on 
the deduced emittance goes from positive values for 
large angles to negative values for small angles. 
\subsection{Active sweep range of scanner}
\begin{figure}
\begin{center}
\epsfig{file=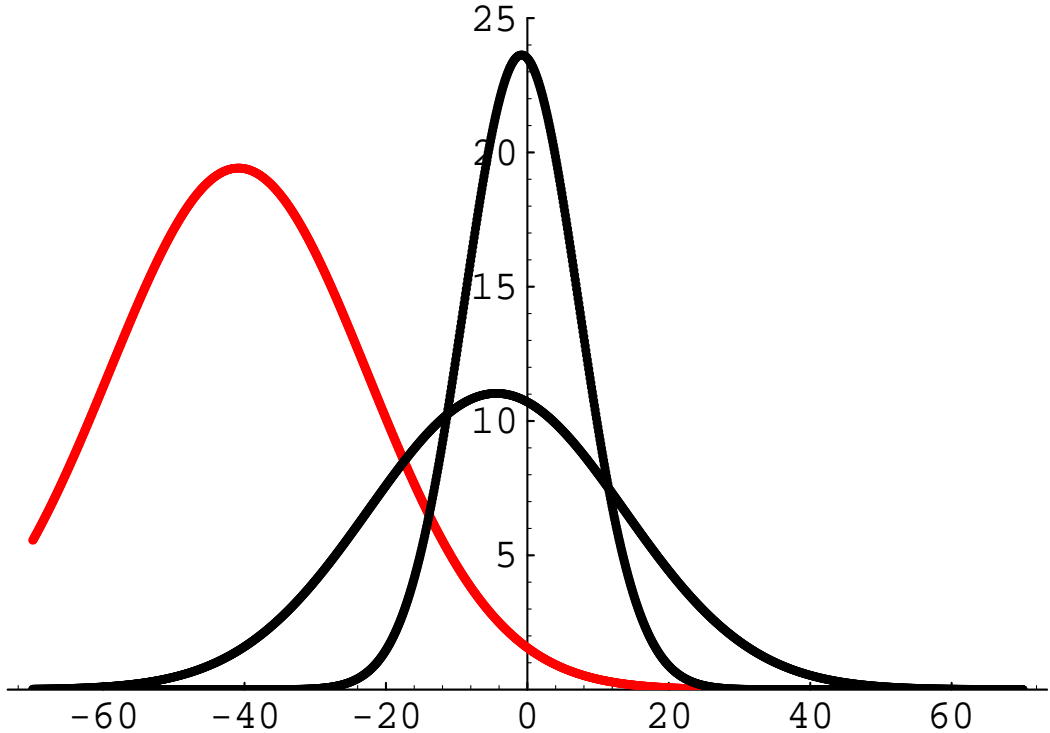,height=10cm,width=14cm,clip=}
  \caption{In the centre of the figure the measured profiles for two 
           beams with the same normalised
           emittance at two different energies,
           100 MeV and 1.4 GeV, are shown. The wider profile at 100 MeV is
	   slightly deformed due to geometrical effects. To the left
	   in the figure a profile truncated at $2\sigma$ from the profile
	   centre is shown. The wire scanner software will for this profile
	   deduce an emittance which is 20\% smaller than the 
	   true emittance.}
\label{f:profile}
\end{center}
\end{figure}

The large transverse size of the large emittance beam demand a long active
sweep range for the scanner to $i$) establish the zero baseline and 
$ii$) avoid acquisition of ``cut'' profiles.
At the CERN-CPS we have measured systematic 
differences of up to 10\% between different wire scanners measuring the
same beam but at different positions with different centres of the closed 
orbit. The numerical "simulations" discussed in the previous section confirms
that such large deviations from the original beam emittance easily can
be caused by a large offset of the centre of the profile. In \fig{f:profile}
the measured beam profile at 100 MeV for a beam profile well centred in 
the wire-scanner sweep range and for one truncated at $2\sigma$ form the 
beam profile centre 
are shown. The presently used wire scanner software will for this 
profile deduce
an emittance which is 20\% smaller than the true beam emittance.  
%
%
\section{Discussion}
We have presented new simulations for the creation
of secondary particles in a thin carbon wire by a primary
proton beam. The derived limits for the use of wire scanners 
show that the use of these devices 
in a low energy (50 MeV - 1 GeV) accelerators with large transverse 
beam size is fully feasible. The total
deposited energy in the wire increases with decreasing beam energy.
Nevertheless, the wire will not get hotter but rather the opposite due 
to the increase of the total heated wire volume as the beam size is usually
large at lower energies. The fact that a large beam can not be considered
as a point source in relation to the detectors will only have small,
and for most measurements, insignificant effect. The emittance blow-up
will increase at lower energies but will for most practical purposes 
be of little importance. However, the large beam size requires a long active
sweep range of the wire scanners to avoid cut-off effects which can result
in significant deviations of the measured emittance from the true beam 
emittance.  
%
\section{Acknowledgements}
Many thanks to Dr.~Charles Steinbach for helpful discussions and to Marco
Pullia for patiently checking our equations and for helpful comments.

\end{document}